\documentclass[11pt]{article}
\usepackage[a4paper]{geometry}
\usepackage[utf8]{inputenc}
\usepackage{amsmath,amssymb}
\usepackage{bm}
\usepackage{cite}
\usepackage{fullpage}
\usepackage{graphicx}
\usepackage{epstopdf}
\usepackage[utf8]{inputenc}
\usepackage[english]{babel}
\usepackage{csquotes}
\usepackage{comment}
\usepackage[dvipsnames]{xcolor}
\usepackage{float}
\usepackage{caption}
\usepackage{subcaption}
\usepackage{comment}
\usepackage{authblk}
\allowdisplaybreaks
\newcommand{\av}[1]{\left< #1 \right>}
\DeclareMathOperator{\Tr}{Tr}
\usepackage{mathtools}
\DeclarePairedDelimiter\bra{\langle}{\rvert}
\DeclarePairedDelimiter\ket{\lvert}{\rangle}
\DeclarePairedDelimiterX\braket[2]{\langle}{\rangle}{#1 \delimsize\vert #2}

\title{Thermodynamics of one and two-qubit nonequilibrium heat engines running between squeezed thermal reservoirs}

\author[1]{Ashutosh Kumar}
\author[2]{Trilochan Bagarti}
\author[1]{Sourabh Lahiri}
\author[3]{Subhashish Banerjee}

\affil[1]{Birla Institute of Technology Mesra, Ranchi, Jharkhand 835215, India}
\affil[2]{Graphene Center, Tata Steel Limited, Jamshedpur-831007, India}
\affil[3]{Indian Institute of Technology Jodhpur-342030, India}

\date{}

\begin{document}

\maketitle

\begin{abstract}
Quantum heat engines form an active field of research due to their potential applications. There are several phenomena that are unique to the quantum regime, some of which are known to give these engines an edge over their classical counterparts. In this work, we focus on the study of  one and two-qubit finite-time Otto engines interacting with squeezed thermal baths, and discuss their important distinctions as well as the advantage of using the two-qubit engine. In particular, the two-qubit engine offers an interesting study of the interplay between the degree of squeezing and that of the coherence between the two qubits.   We find that the two-qubit engine generally yields higher power than its one-qubit counterpart. The effective temperature of the squeezed baths can be calculated both for the one and two-qubit engines, and they tend to show an exponential growth with increase in squeezing parameters $r_h$ and $r_c$. It is also observed that by tuning the squeezing parameters, the machine can be made to work either in the engine or in the refrigerator mode. Additional effects due to the change in the inter-qubit separation have been studied.
\end{abstract}

\section{Introduction}

The topic of thermal machines at small scales has attracted a lot of attention, especially in the last decade \cite{roldan2016,bechinger2012,lahiri2020,saha2019,Aradhana2021}. Not only do their study help in enhancing our understanding of the thermodynamics for small systems, they are also of immense practical importance and have been predicted to potentially revolutionize the medical industry by powering tiny machines. A discussion on the impact of such ``nanobots'' on healthcare is provided in \cite{Sorna2019,saadeh2014nanorobotic}. The protocol for using a Brownian particle as the working system of an engine was suggested in \cite{sei08_epl}. One of the first experiments incorporating such concepts to construct a Stirling engine was in \cite{bechinger2012}. A mechanical autonomous engine was obtained experimentally \cite{Garcia2016}. Experimental realization of heat engines in an active or non-thermal bath was investigated in \cite{ajay2016_nature}. These works were subsequently investigated theoretically \cite{lahiri2020,saha2019,Aradhana2021}. 

Parallelly, the studies in quantum heat engines/refrigerators have been another ramification in this area \cite{quan2007,campisi2014,alicki1979,joh18_pre,Subhashish2018,Jiao2021,Lutz2020,kosloff2013quantum,vinjanampathy2016quantum}. The field was pioneered by the seminal work in \cite{Scovil1959}, where a three-level maser had been used to undergo an engine protocol. There have been several experimental demonstrations of quantum thermal engines \cite{Peterson2019,Abah2016,Anders2019}. A number of works have addressed the effect of entanglement \cite{zhang2007four} and coherence \cite{scully2003extracting,Scully2007,Scully2011,Abah2014} on the efficiency of the engine. Some recent works explore the effect of squeezed thermal baths \cite{subhashish2019open,Ficek2002}  on the engine performance  \cite{Zhang2020,Xiao2022,Mohanta2022}.
In \cite{Subhashish2018}, it was shown that use of non-Markovian heat baths apparently seem to violate the thermodynamic Second Law, whose resolution lies in redefining work and heat from physical considerations. In \cite{Zhang2020}, under the conditions of quantum adiabaticity (quasistatic process), the dependence of efficiency of an Otto engine at maximum power of a harmonic oscillator with varying stiffness on the squeezing parameters was investigated. 

Here we study the single qubit Otto engine in a nonequilibrium setup (finite cycle time) in presence of squeezed thermal reservoirs, and then extend it to a system consisting of two qubits, making use of the treatment in \cite{Banerjee2010b}. This entails the use of master equations, well known from the field of Open Quantum Systems \cite{subhashish2019open,Breuer2002}. The master equations are worked out  numerically, and their outputs are compared. We first show the dependence of the engine power on the squeezing parameters, both for the one-qubit and the two-qubit systems. 

The plan of the work is as follows. In section \ref{sec:model}, we provide an introduction to the engine model studied, including the one-qubit and two-qubit working mediums, along with the details of the protocol as well as the master equations used to take into account the interaction with the baths. In section \ref{sec:thermodynamics} we discuss the list of thermodynamic quantities to be evaluated. These will then be followed up with a discussion of the workings of the one-qubit and two-qubit engines in sections \ref{sec:OQE_results} and \ref{sec:TQE_results} respectively, under ambient conditions. We finally make our conclusions in section \ref{sec:conclusions}.

\section{The Model} \label{sec:model}
\begin{figure}[h!]
	\begin{center}
	\includegraphics[width=0.8\textwidth]{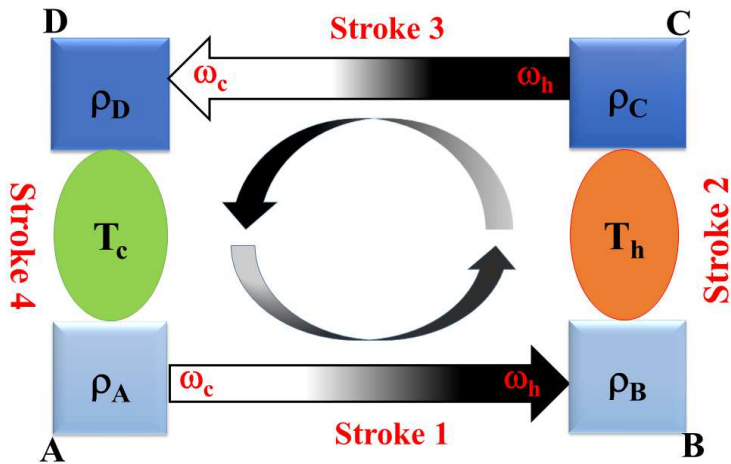}
	\caption{Schematic diagram of the quantum Otto cycle with squeezed thermal reservoirs.}
		\label{fig:SchematicOtto}
	\setlength{\belowcaptionskip}{1pt}
	\end{center}

\end{figure}

The traditional Otto cycle has two isochoric (constant volume) and two adiabatic operations and employs an ideal gas as the working medium. 
During the isochoric processes, heat is exchanged with the thermal reservoirs, whereas work is done during adiabatic processes. 
In the quantum model of an Otto cycle, the working medium is a quantum system, such as a spin-1/2 system \cite{quan2007}. The quantum isochoric process 
involving a two-level system keeps the energy-level spacing unchanged (instead of fixing the volume as in a classical system). The system being connected to a reservoir during the isochoric phase, the evolution is nonunitary in these steps, whereas it remains unitary in the adiabatic processes. The schematic diagram of a quantum Otto cycle has been given in Fig. \ref{fig:SchematicOtto}. The full cycle is given by the process  A$\rightarrow$B$\rightarrow$C$\rightarrow$D$\rightarrow$A, labeled as Strokes 1, 2, 3 and 4, respectively.

\subsection{Model for One-Qubit Engine (OQE)} \label{subsec:OQE_model}

    We now consider the evolution of the density operator, described by the von Neumann equations in the unitary steps A$\to$B and C$\to$D, and by the Lindblad equations during the dissipative steps B$\to$C and D$\to$A (see fig. \ref{fig:SchematicOtto}). The density operator is a $2\times 2$ matrix, which in the energy basis $\{|g\rangle,|e\rangle\}$ is given by $\begin{pmatrix}
     \rho_{ee} & \rho_{eg}\\ \rho_{ge} & \rho_{gg}    \end{pmatrix}$. The eigenvectors $|g\rangle$ and $|e\rangle$ correspond to the ground and the excited states, respectively.

	\paragraph{Stroke 1:} We consider a  quantum Otto cycle for a time-dependent Hamiltonian. The frequency of the Hamiltonian is linearly modulated with time from $\omega{_c}$ to $\omega{_h}>\omega_c$. The system Hamiltonian changes from $H(\omega_c) $ to $H(\omega_h)$. The density matrix $\rho(t)$ is evolved using the von-Neumann equation

\begin{align}
\frac{\partial \rho(t)}{\partial t} &=-i[H(\omega(t)),\rho(t)],
\end{align}

where $H(\omega(t)) =(\omega(t)/2)\sigma_z$, and
\begin{align}
\omega(t) &=\omega_c(1-t/\tau)+\omega_ht/\tau.
\label{eq:H_AB}
\end{align}

In the above equations, $\sigma_z$ is the Pauli matrix given by $\begin{pmatrix}
    1 & 0 \\
    0 & -1
\end{pmatrix}$.

Throughout this article, we set the Planck's and the Boltzmann's constant to unity (i.e., $\hbar=1,~ k_B = 1$) for convenience. The time duration of each stroke is set to a finite time interval $\tau$, which gives a total cycle time of $4\tau$, and makes the process a non-equilibrium one.

\paragraph{Stroke 2:} In this stroke, the hot bath at temperature $T_h$ is connected to the system. The energy spacing is held constant at $\omega_h$, resulting in $H(\omega_h)$ =$(\omega_h/2)\sigma_z$ as the final Hamiltonian at point C. 
Heat is transferred from the hot bath, modeled here as a squeezed thermal bath, to the working medium during this process.

    The density matrix $\rho(t)$  of the system evolves through the quantum master equation as given below, where we ignore the Lamb shift contribution to the Hamiltonian \cite{subhashish2019open,Srikanth2008,Omkar2013}:
 \begin{equation}
	\begin{split}
		\frac{\partial \rho(t)}{\partial t} =-i[H(\omega_h),\rho(t)]+\sum_{j=1,2}\left[2R_j\rho(t) R^\dagger_j - \left\{R^\dagger_j R_j,\rho(t)\right\} \right],
	\end{split}
	\label{eq:master}
\end{equation}
where $R_1=\left[\sqrt{\gamma_0(N_{th}+1)/2}\right]R$, $R_2=\left[\sqrt{\gamma_0 N_{th}/2}\right]R^\dagger$, $N_{th}=1/(e^{\beta_h\omega_h}-1)$ is the Planck distribution,  $R=\cosh(r)\sigma_{-}+\exp(i\phi)\sigma_{+}\sinh(r)$, $r$ and $\phi$ are squeezing parameters. 
$\sigma^+$ and $\sigma^-$ are the raising and lowering operators: $\sigma^+ = |e\rangle\langle g|$, $\sigma^- = |g\rangle\langle e|$.
The notation $\{A,B\}$ implies the anti-commutator of operators $A$ and $B$.
Note that $N_{th}$ gives the average number of photons emitted with frequency $\omega_h$ and inverse temperature $\beta_h \equiv 1/T_h$, and $\gamma_0$ is the spontaneous emission rate.

\paragraph{Stroke 3:} The system is then decoupled from the bath. 
The frequency corresponding to the energy level spacing is changed linearly from $\omega{_h}$ to $\omega{_c}$.

The density matrix $\rho(t)$ is evolved using the von-Neumann equation
\begin{equation}
	\frac{\partial \rho{(t)}}{\partial t} = -i[H(\omega(t)),\rho{(t)}],
\end{equation}
where $H(\omega(t))=(\omega(t)/2)\sigma_z$, with
\begin{align}
    \omega(t)=\omega_c(t/\tau-2)+\omega_h(3-t/\tau).
    \label{eq:H_CD}
\end{align}

\paragraph{Stroke 4:} The system is now connected to the cold bath. The energy gap is held constant at  $\omega_c$, resulting in $H(\omega_c)=(\omega_c/2)\sigma_z$ as the final Hamiltonian at the end of this stroke. The master equation is of the same form as in Eq. \eqref{eq:master}, except that the Hamiltonian appearing in the commutator is now $H(\omega_c)$.


\subsection{Model for Two-Qubit Engine (TQE)} \label{subsec:TQE_model}
\label{subsec:two-qubit}

Here, in the isochoric strokes of the engine, we use a two-qubit system as the working medium interacting with squeezed thermal bath, with a position dependent system-bath coupling \cite{Banerjee2010a,Banerjee2010b,Ficek2002}. This enables the dynamics to be considered in the independent and collective regimes. 
The quantum master equation for the two-qubit system, with the energy spacing between the independent qubits (when they are not interacting) being $\omega_1$ and $\omega_2$ respectively,  in contact with the  reservoir taken to be in the squeezed thermal state, is:
 \begin{align}
   \frac{\partial \rho(t)}{\partial t} =
	- i[\tilde{H},\rho(t)] &-\frac{1}{2}\Gamma_{12}\sum_{i,j=1,2}[1+\Tilde{N}](\rho S_i^+S_j^- +S_i^+S_j^-\rho -2S_j^-\rho S_i^+) \nonumber\\
	&-\frac{1}{2}\Gamma_{12}\sum_{i,j=1,2}\Tilde{N}(\rho S_i^-S_j^+ +S_i^-S_j^+\rho -2S_j^+\rho S_i^-) \nonumber\\
	&+\frac{1}{2}\Gamma_{12}\sum_{i,j=1,2}\Tilde{M}(\rho S_i^+S_j^+ +S_i^+S_j^+\rho -2S_j^+\rho S_i^+) \nonumber\\
	&+\frac{1}{2}\Gamma_{12}\sum_{i,j=1,2}\Tilde{M^*}(\rho S_i^-S_j^- +S_i^-S_j^-\rho -2S_j^-\rho S_i^-),
	\label{eq:two_master}
     \end{align}
  where
 \begin{align}
   \tilde{N} &=N_{th}[\cosh^2(r)+\sinh^2(r)]+\sinh^2(r); \nonumber\\
 \tilde{M} &=-\frac{1}{2}\sinh(2r)\exp{(i\phi)}(2N_{th}+1);\nonumber\\
    N_{th} &=\frac{1}{\exp(\hbar\omega_0/k_BT)-1}; \hspace{1cm}
 \omega_0  =\frac{\omega_1+\omega_2}{2}.
   \end{align}
  Here, $\Tilde{M}^*$ is the conjugate of $\Tilde{M}$, $N_{th}$ is the Planck distribution function giving the number of thermal photons at the frequency $\omega_0$. We have defined $\omega_1 = \omega_{1h}$ and $\omega_2 = \omega_{2h}$  as the energy-level spacings of qubits 1 and 2, respectively, that occur in stroke 2 when each qubit hypothetically undergoes an OQE (independently of the other qubit). Similarly, $\omega_1 = \omega_{1c}$ and $\omega_2 = \omega_{2c}$ are the energy-level spacings when the corresponding qubits are in contact with the cold bath, in absence of the other qubit. Note that when both qubits are present, as in the case of the TQE, the parameters $\omega_1$ and $\omega_2$ get subsumed into $\omega_0$, which becomes the appropriate parameter to look at.
  At cold bath temperature $T_c$, we have $\omega_0 \equiv \omega_{0c}=(\omega_{1c} + \omega_{2c})/2$, and at hot bath temperature $T_h$,  $\omega_0 \equiv \omega_{0h}=(\omega_{1h} + \omega_{2h})/2 $. The  $r$ and $\phi$ are the squeezing parameters and $\Gamma_{12}$ is the collective spontaneous emission rate, given in terms of the individual spontaneous emission rates $\Gamma_1$ and $\Gamma_2$ by 
  \begin{align}
      \Gamma_{12}=\Gamma_{21} = \sqrt{\Gamma_1 \Gamma_2}~F(k_0 r_{12}),
  \end{align}
   where 
   \begin{align}
        \Gamma_i &=\frac{\omega_i^3\mu_i^2}{3\pi\epsilon\hbar c^3}, \hspace{0.5cm}i=1,2, \hspace{1cm}\mbox{and}\nonumber\\
         F(k_0r_{12}) &=\frac{3}{2}\left[\{1-(\bm{\hat{\mu}.\hat{r}}_{12})^2\}
  \frac{\sin{(k_0 r_{12})}}{(k_0 r_{12})}\right. \nonumber\\
  &\hspace{1cm}\left.+\{1-3(\bm{\hat{\mu}.\hat{r}}_{12})^2\}\times
  \left\{\frac{\cos{(k_0 r_{12})}}{(k_0 r_{12})}^2-\frac{\sin{(k_0 r_{12})}}{(k_0 r_{12})^3}\right\}\right].
   \end{align}
   Here, $r_{12}=|\bm{r}_{12}|=r_{21}$ is the magnitude of the position vector  of one qubit with respect to the other, i.e., the distance between the two qubits. Furthermore, we consider identical qubits, for which $\Gamma_1=\Gamma_2\equiv\Gamma$, and atomic transition dipole moments are given by $\bm{\mu}_1=\bm{\mu}_2=\bm{\mu}$.
  The hat symbols imply unit vectors in the above equation.

   A similar equation can be written for the dynamics of the system when it is in contact with the cold bath.
  Note that in absence of squeezing, $\tilde M=0$, and the last two terms on the right hand side of Eq. \eqref{eq:two_master}  would disappear.
  The equations for the unitary processes $A\to B$ and $C\to D$ are given by von Neumann equations, as in the case of the OQE. 
 We change the energy gap corresponding to  $\omega_0$ with time linearly during the unitary steps, with the same functional forms as in the case of OQE, see Eqs. \eqref{eq:H_AB} and \eqref{eq:H_CD}.

 The Hamiltonian appearing in the {\it RHS} of Eq. (\ref{eq:two_master}), consists of a total energy operator for the two qubits, and an interaction between them, which is mediated by the qubits' interaction with the bath. It is given by:
\begin{equation}
  \tilde{H} = \hbar (\omega_1 S^{\rm z}_1 + \omega_2 S^{\rm^ z}_2)+
  \hbar \Omega_{12}(S^{\rm +}_1 S^{\rm -}_2 + S^{\rm +}_2 S^{\rm -}_1),\\
  \label{eq:two_hamiltonian}
 \end{equation}
where 
  \begin{align}
S^{\rm z}_1  &= \frac{1}{2}(\ket{e_1}\bra{e_1}-\ket{g_1}\bra{g_1}), \hspace{0.5cm}
S^{\rm z}_2  = \frac{1}{2}(\ket{e_2}\bra{e_2}-\ket{g_2}\bra{g_2})\nonumber\\
S^{\rm +}_1  &= \ket{e_1}\bra{g_1},\hspace{0.2cm}  S^{\rm +}_2 = \ket{g_2}\bra{e_2}, \hspace{0.2cm}
S^{\rm -}_1  = \ket{e_1}\bra{g_1},\hspace{0.2cm}  S^{\rm -}_2 = \ket{g_2}\bra{e_2}.\nonumber\\   
  \end{align}
Here $S^{\pm}_{1,2}$ are the raising and lowering operators and $S^{\rm z}_{1,2}$  are the energy operators of the concerned qubits.
 The strength of the dipole-dipole interaction between the two qubits is given by
 
  \begin{align}
  \Omega_{12}=\Omega_{21} &=\frac{3}{4}\Gamma\left[-\{1-(\bm{\hat{\mu}.\hat{r}}_{12})^2\}
  \frac{\cos{(k_0 r_{12})}}{(k_0 r_{12})}\right. \nonumber\\
  &\left.+\{1-3(\bm{\hat{\mu}.\hat{r}}_{12})^2\}\times
  \left\{\frac{\sin{(k_0 r_{12})}}{(k_0 r_{12})}^2+\frac{\cos{(k_0 r_{12})}}{(k_0 r_{12})^3}\right\}\right].
  \label{eq:OmegaGamma}
  \end{align}
  The magnitude of the wavevector is given by $k_0=2\pi/\lambda_0 = \omega_0/c$, where $\lambda_0$ is the resonant wavelength.
The decoherence may be broadly categorized as (a) independent decoherence where $\bm{k_0.r}_{12}> 1$ and (b) collective decoherence where $\bm{k_0.r}_{12}\ll 1$.  
$\Omega_{12}$ provides the shifts in atomic energy levels.
  
The dressed state basis is found convenient to analyse the collective two-qubit dynamics interacting with a squeezed thermal bath. It is given by:
\begin{align}
      \ket{g} &= \ket{g_1}\ket{g_2},\nonumber\\
      \ket{s} &= \frac{1}{\sqrt{2}}(\ket{e_1}\ket{g_2} + \ket{g_1}\ket{e_2}),\nonumber\\
      \ket{a} &=\frac{1}{\sqrt{2}}(\ket{e_1}\ket{g_2} - \ket{g_1}\ket{e_2}),\nonumber\\
      \ket{e} &= \ket{e_1}\ket{e_2},\nonumber\\
      \label{eq:dressed_basis}
\end{align}
 with the corresponding eigenvalues being $E_g = -\hbar\omega_0$, $E_s = \hbar\Omega_{12}$, $E_a = -\hbar\Omega_{12}$ and $E_e = \hbar\omega_0$ (it may be recalled that a change in basis does not affect the eigenvalues).
 The master equation, Eq. \eqref{eq:two_master}, can be solved to yield the reduced density matrix   taken in the two-qubit dressed state basis, Eq. \eqref{eq:dressed_basis}. The various density matrix elements are provided in the Appendix \ref{app:twoQubit}.
In this basis, the density matrix is given by
 \begin{align}
     \rho &= \begin{pmatrix}
      \rho_{ee} & \rho_{es} & \rho_{ea} & \rho_{eg}\\
      \rho^*_{es} & \rho_{ss} & \rho_{sa} & \rho_{sg}\\
      \rho^*_{ea} & \rho^*_{sa} & \rho_{aa} & \rho_{ag} \\
      \rho^*_{eg} & \rho^*_{sg} & \rho^*_{ag} & \rho_{gg}
     \end{pmatrix}.
 \end{align}
The equation of motion for each element of $\rho$ in the dressed state basis and the subsequent transformation of $\rho$ to the basis $\{\ket{g_1},\ket{g_2},\ket{e_1},\ket{e_2}\}$ have been discussed in appendix \ref{app:twoQubit}.

\section{Calculation of thermodynamic quantities} \label{sec:thermodynamics}

The important thermodynamic quantities that we are interested in are listed below:

\begin{enumerate}
    \item {\bf Work:} The output work is of fundamental importance in the study of any heat engine. In the Otto engine, it is simple to define work and heat, since the steps in the cycle where work is extracted are different from the steps where heat is exchanged. The unitary processes ($A \to B$ and $C \to D$) constitute the steps where work is done/extracted. The works  \textit{extracted} (it may be noted that the work done and work extracted are related simply by the switching of signs) from the system in these two steps are defined mathematically as:
    \begin{align}
        \av{W_{AB}} &= \av{E_A}-\av{E_B};\nonumber\\
        \av{W_{CD}} &= \av{E_C} - \av{E_D}.
    \end{align}
   Here, the average energy at state A is $\av{E_A}\equiv \Tr[\rho_A H_A]$, $\rho_A$ being the reduced density matrix and $H_A$ being the Hamiltonian operator at the point $A$. The latter is given by $H_A = H(\omega_c)$. The other energy averages are taken in a similar manner.

   \item {\bf Power:} 
   	   Power is obtained by dividing the total work done by the cycle time.

   \item {\bf Heat:} The heats \textit{absorbed} by the system are defined as:
   \begin{align}
       \av{Q_{BC}} = \av{E_C}-\av{E_B};\nonumber\\
       \av{Q_{DA}} = \av{E_A}-\av{E_D}.
   \end{align}
   
   Note that  $\av{Q_{DA}}$ will be negative, since heat is released into the cold reservoir in this step.

   \item {\bf Efficiency:} The efficiency is defined as the ratio of work extracted in the full cycle to the heat absorbed from the hot bath:
   \begin{align}
       \eta \equiv \frac{\av{W_{AB}}+\av{W_{CD}}}{\av{Q_{BC}}}.
   \end{align}

  \item {\bf Efficiency at maximum power:} Here, the power is maximized with respect to a given system parameter, and the value of efficiency corresponding to this particular value of the parameter gives the efficiency at maximum power. As shown in \cite{cur75_ajp}, for not too large temperature difference between the two heat baths, the value of this efficiency is given by the Curzon-Ahlborn form
  \begin{align}
      \eta_{CA} = 1-\sqrt{\frac{T_c}{T_h}}.
      \label{eq:EtaCA}
  \end{align}
\end{enumerate}

In the next two sections, we numerically calculate the thermodynamic observables for the OQEs and TQEs, respectively.

\section{Results and Discussions on OQE} \label{sec:OQE_results}
\label{sec:OQE}

Throughout the current and the next section, we set the squeezing parameters $\phi_c$ and $\phi_h$ to zero, since none of the thermodynamic observables that we have investigated are found to depend on these parameters. Furthermore, the duration of each stroke has been set to $\tau=2$ throughout this work. The rate of spontaneous emission $\gamma_0$ has been taken to be equal to 1 for all the outputs generated for the OQE.

\begin{figure}[h!]
\centering
	\includegraphics[width=0.55\textwidth]{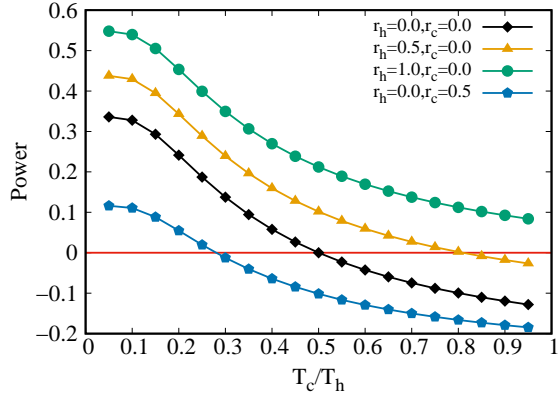}
\caption{Variation of power of an OQE as a function of the temperature ratio $T_c/T_h$ for different squeezed parameters $r_h$ and $r_c$. Other parameters are $\omega_h=20$, $\omega_c=10$. }
\label{fig:PvsT}
\end{figure}

\paragraph{Power of the engine:} In fig. \ref{fig:PvsT}, the variation of power as a function of the ratio of the temperatures of the cold and hot baths, $T_c/T_h$, is shown. It is observed that as the temperature difference decreases, i.e., as $T_c/T_h$ gets closer to unity, the value of power decreases, in conformity with our expectations. The different curves are for different bath squeezing parameters. The red horizontal line gives the locus of zero power. The third curve from the top (black line and solid squares) is the reference curve, where there is no squeezing ($r_h=r_c=0$). We find that if $r_h$ increases relative to $r_c$, the power output of the engine increases, thus indicating that the system behaves like one that is connected to reservoirs with higher temperature difference. In contrast, in the opposite regime as denote by the lowest curve ($r_h=0,~r_c=0.5$, cyan line with solid pentagons), the engine performs worse than the reference case.
Summarizing, we notice that for a given set of bath temperatures, the value of power increases with increase in $r_h$, while it decreases with increase in $r_c$. This might be an indicator that the engine experiences an effective temperature that increases with the increase in the squeezing parameter $r$.

\begin{figure}[h!]
\centering
    \begin{subfigure}{0.45\textwidth}
        \includegraphics[width=\textwidth]{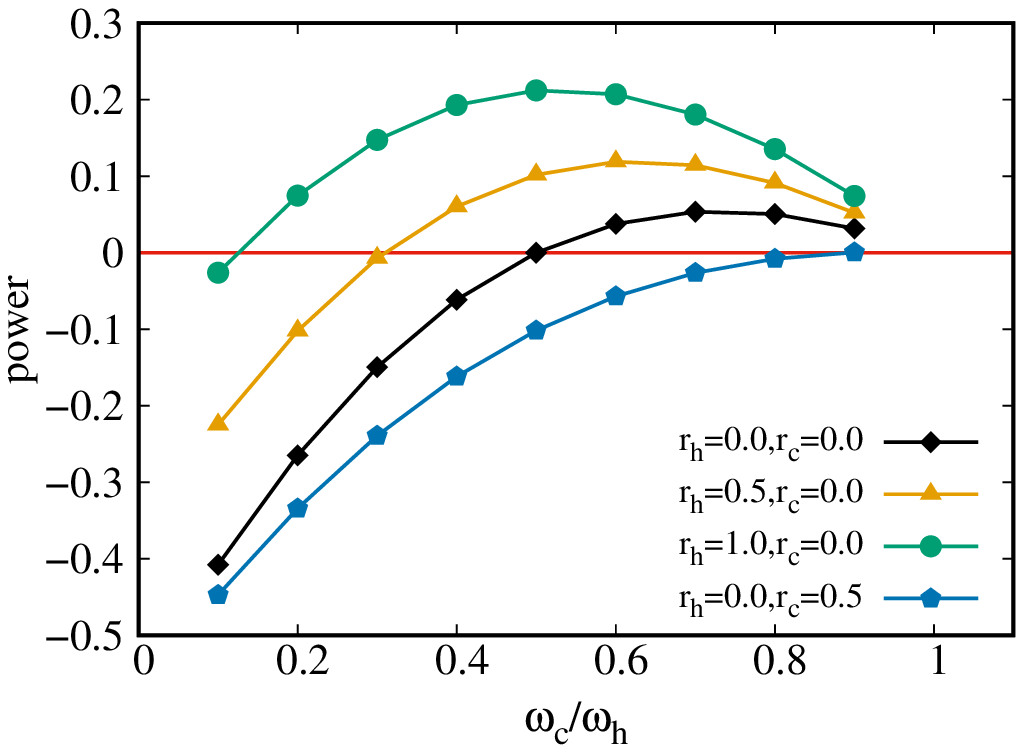}
        \caption{}
    \end{subfigure}
    \begin{subfigure}{0.45\textwidth}
    \centering
        \includegraphics[width=\textwidth]{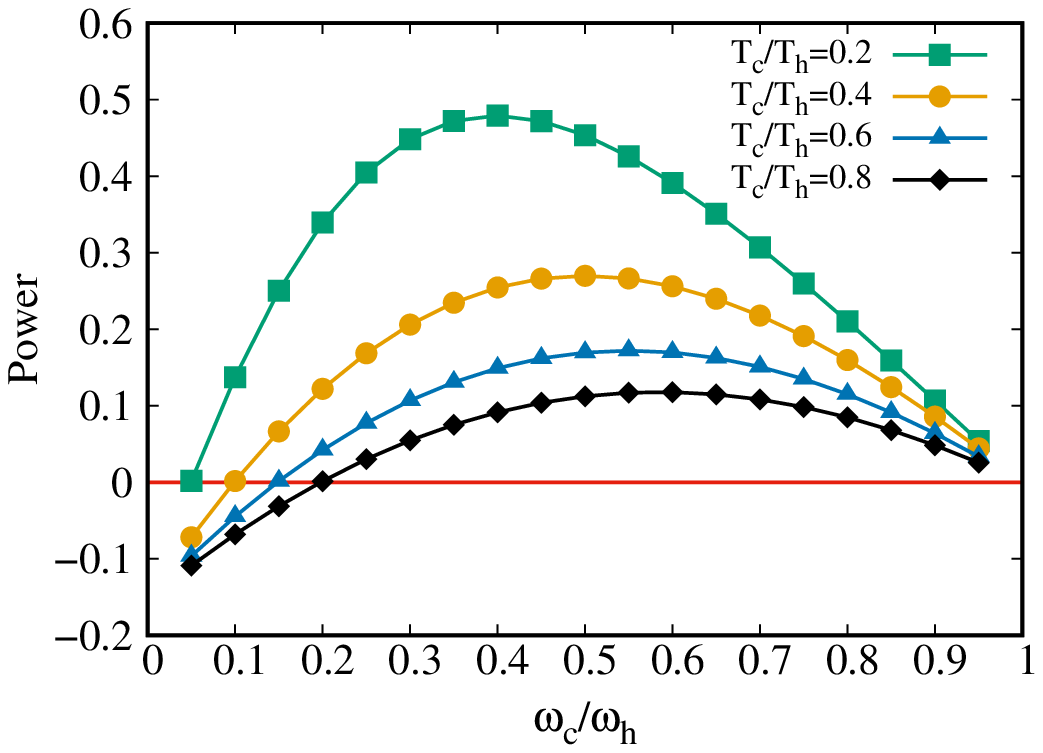}
        \caption{}
    \end{subfigure}
	\caption{(a) Power of an OQE as a function of $\omega_c/\omega_h$ for different squeezing parameters. Other parameters are $T_h= 20, ~T_c= 10$. (b) Power as a function of  $\omega_c/\omega_h$ for different value of bath temperature ratio, with $r_h=1,~r_c=0$. Others parameter are same as in (a).}
		\label{fig:PvsOmega}
	\setlength{\belowcaptionskip}{1pt}
\end{figure}

 In fig. \ref{fig:PvsOmega}(a), the functional dependence of power on the ratio $\omega_c/\omega_h$ has been shown for different sets of values of $r_c$ and $r_h$ (the reader may recall that $\omega_h$ is the energy difference between the two levels when in contact with the hot bath, while $\omega_c$ is that when in contact with the cold bath). The other parameters are kept fixed, with their values as mentioned in the figure caption. A clear non-monotonicity in the variation of power is noted, except in the case where $r_c>r_h$, where the system mostly consumes work and thus fails to act as an engine (fourth curve from the top, solid petagons). 
 In fig. \ref{fig:PvsOmega}(b), similar dependence has been shown, but this time the values of the squeezing parameters have been held fixed (see caption),  while different temperature ratios have been used to obtain the different curves. As the difference between bath temperatures increases, i.e., $T_c/T_h$ decreases, the output power increases as well, in agreement with our expectations.

\paragraph{Efficiency at maximum power:} We next investigate the dependence of efficiency at maximum power ($\eta_{MP}$) on the ratio of temperatures $T_c/T_h$. 
By recording the values of $T_c/T_h$ at the maxima of the curves in fig. \ref{fig:PvsOmega}(b), one can obtain the variation of the efficiency at maximum power $\eta_{MP}$ with the ratio of temperatures, where the latter can be calculated using
\begin{align}
    \eta_{MP} \equiv 1 - \left(\frac{\omega_c}{\omega_h}\right)_{\rm peak},
     \label{eq:Emax_p}
\end{align}
 the second term on the right hand side being the ratio of frequencies in the state where the power shows a peak \cite{Das2020}. The derivation of the equation \eqref{eq:Emax_p} has been provided in appendix \ref{app:eta}.

\begin{figure}[h!]
\centering
	\includegraphics[width=0.55\textwidth]{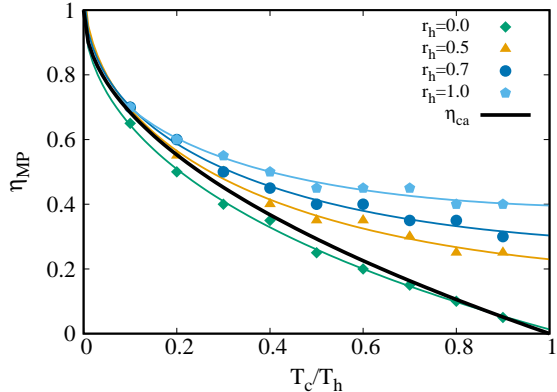}\\
	\caption{$\eta_{MP}$ of an OQE as a function of $T_c/T_h$ for different squeezed parameters $r_h$, with $r_c=0$.}
	\label{fig:EtaMPvsTemp}
\end{figure}

By changing the squeezing parameters (here we change $r_h$, keeping $r_c=0$), a family of such curves can be generated, some of which we have shown in fig. \ref{fig:EtaMPvsTemp}. 
The solid curves represent the best fits of the data points, where small errors are induced due to inaccurate extraction of the aforementioned maxima.
We observe that the curve for the thermal bath, corresponding to $r_h=0$, as shown by the lowest curve (green line with solid rhombuses) is in good agreement with the functional form given by Eq. \eqref{eq:EtaCA}, corresponding to the curve labelled by $\eta_{CA}$ (black line without symbols), when the difference between the temperatures of the two baths is small enough, i.e., when $T_c\to T_h$ (linear response regime).
As the value of $r_h$ increases, the deviations from the functional form of Eq.  \eqref{eq:EtaCA} clearly increases. The fits were obtained using the functional form $y(x)=ax+b\sqrt{x}+c$, as is expected from the expansion of $\eta_{MP}$ in powers of the Carnot efficiency $\eta_C$ (see \cite{Bassie2022,sei08_epl}) via the optimization of the constants $a$, $b$ and $c$.
%


\paragraph{Effective temperatures:} An interesting feature of systems interacting with squeezed thermal baths is that they can be described by means of effective temperatures, as shown in \cite{Klaers2017,Manzano2018,Zhang2020}. 
However, we have a two-state system in a nonequilibrium steady state (it is to be noted that in absence of squeezing, the cycle period of the engine is more than sufficient to make the process an equilibrium one). It would be interesting to check whether the form of the effective temperature is the same as in the case of a harmonically bound system at equilibrium with the heat bath.

\begin{figure}[h!]
 \centering
 	\begin{subfigure}{0.48\textwidth}
	\includegraphics[width=\textwidth]{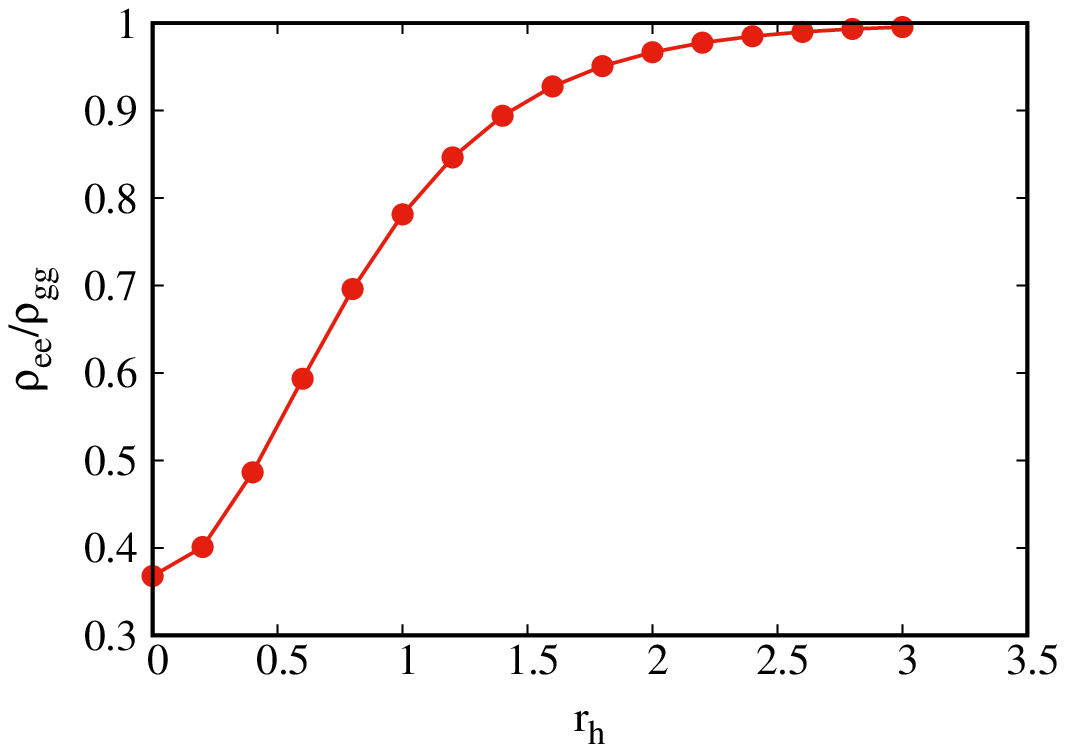}
	\caption{}
\end{subfigure}
 \begin{subfigure}{0.48\textwidth}
     \centering
     \includegraphics[width=\textwidth]{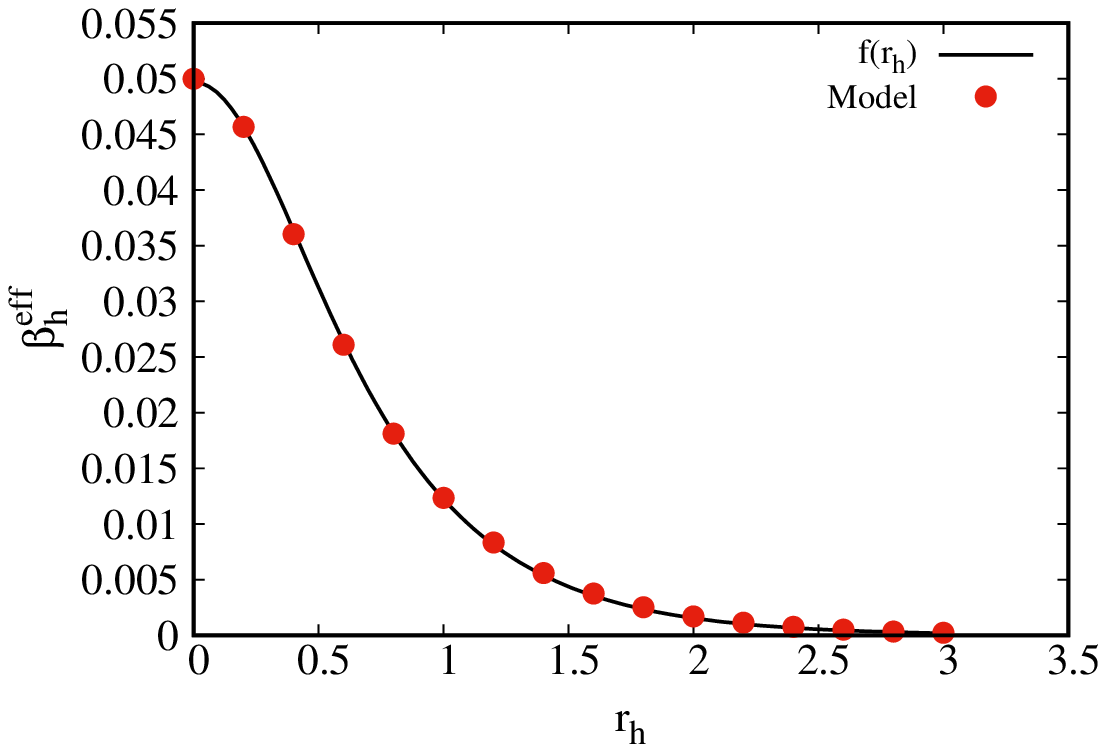}
     \caption{}
 \end{subfigure}
 \caption{(a)  Ratio of $\rho_{ee}$ and $\rho_{gg}$ as a function of $r_h$ for BC Stroke.
  (b) Variation of effective inverse temperature as the function of squeezing parameter. The parameters used are $\omega_h=20$, $\omega_c=10$, $T_h=20$, $T_c=10$. The black solid line is given by the function $f(r_h)=a ~\mbox{sech}(b r_h)$, where $a=0.0496, ~b=2.079$.}
 
    \label{fig:Teff}
\end{figure}

We have studied this dependence in fig. \ref{fig:Teff}. Figure \ref{fig:Teff}(a) shows the variation of the ratio of the populations of the excited and the ground levels, $\rho_{ee}/\rho_{gg}$, as a function of $r_h$, when the system has been allowed to evolve in presence of the hot squeezed thermal reservoir, and is in state C (see fig. \ref{fig:SchematicOtto}). 
Now, an effective inverse temperature $\beta_h^{\rm eff}$ can be defined (a similar method has been adopted in \cite{Xiao2022}), if this function is equated to a Boltzmann factor, $\exp(-\beta_h^{\rm eff}\omega_h)$, as shown in fig. \ref{fig:Teff}(b). One can readily compute the effective temperature $T^{\rm eff}_h$ of the system by using the relation $T^{\rm eff}_h = 1/\beta^{\rm eff}_h$.
The data has been fitted with the function $f(r_h)=a~ {\rm sech}(2r_h)$, and we find the parameters to be $a\simeq 0.05$ and $b\simeq 2$ (the accurate values are provided in the figure caption). This shows that the effective inverse temperature closely follows the relation $\beta_h^{\rm eff} =\beta ~{\rm sech}(2r_h)$ (see \cite{Manzano2018}), where $\beta\equiv 1/T_h$ is the actual inverse temperature of the bath. 
Interestingly, however, we found the coefficients to remain approximately the same ($a=0.0497,~b=2.080$) even when the system is far from a stationary state. Here, it has only partially relaxed in contact with the squeezed heat bath of temperature $T_h$ in stroke 2 after evolving for a very short duration of 0.2, i.e., at net time $t=2.2$, which is closer to the state B than state C (see fig. \ref{fig:SchematicOtto}). This implies that the above relation for inverse effective temperature remains valid to a good accuracy even when the system has been driven out of its stationary state.

\begin{figure}[h!]
     \begin{center}
	\includegraphics[width=0.6\textwidth]{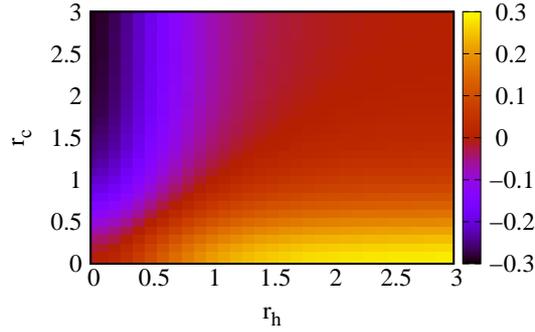}
	\caption{Phase plot for the dependence of the power on the squeezing parameters $r_h$ and $r_c$. The color coding provides the value of power. Other parameter used are $\omega_h=20$, $\omega_c=10$, $T_h=20$, $T_c=10$.}
	\label{fig:denp_1q}
	\setlength{\belowcaptionskip}{1pt}
	\end{center}
    \end{figure}

The phase map between the output power of the OQE and the bath's squeezing settings is shown in fig. \ref{fig:denp_1q}. It should be noted that the machine typically operates in the engine mode when $r_h > r_c $. When $r_c>r_h$, the effective temperature of the cold bath increases much more as compared to that of the hot bath because of the exponential dependence on squeezing parameters, and can even lead to $T^{\rm eff}_c > T^{\rm eff}_h$. This is what typically happens in our setting, as is clear from the phase diagram. Consequently, the squeezing parameters can be tuned to convert the engine into a refrigerator and vice versa. 

\section{Results and Discussions on TQE} \label{sec:TQE_results}
\label{sec:TQE}

We now study the thermodynamics of a two-qubit engine (TQE), whose dynamics has been discussed in Sec. \ref{subsec:two-qubit}, and compare with the behaviour of the corresponding thermodynamic observables of OQE. We show that in general, the TQE yields better output as compared to its single-qubit counterpart. In all the figures in this section, we have set $\phi_h=\phi_c=0,~\Gamma=1,~\tau=2$.

\begin{figure}[h!]
    \centering
        \includegraphics[width=0.5\textwidth]{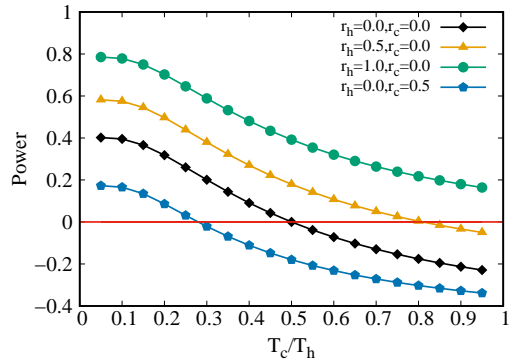}
     \caption{Variation of power of a TQE as a function of the temperature ratio $T_c/T_h$ for different squeezed parameters $r_h$ and $r_c$. Other parameters are: $\omega_{0h}=20$, $\omega_{0c}=10$, $r_{12}$= 0.5.}
      \label{fig:2qeff_PvsT}
    \end{figure}

\paragraph{Power:} Fig. \ref{fig:2qeff_PvsT} depicts the variation of  power, with temperature ratio. 
As mentioned in the figure caption, since the redefined variable  $r_{12}=\bm{k_0.r_{12}}<1$, it implies that the system is undergoing collective decoherence. The rate of spontaneous decay $\Gamma$ is set to 1 just like in the case of the OQE, so as to offer a meaningful comparison between the two.
Just like the OQE, the efficiency for TQE too is dependent only on the frequencies and independent of the squeezing parameters (see appendix \ref{app:eta}).
Also,  the \textit{extracted} power (one that has a positive value) is observed to increase with  $r_h$ but decreases with temperature difference of the baths, in accordance with the qualitative trends already observed for the OQE (see fig. \ref{fig:PvsT}). The former dependence again indicates that an increase in the magnitude of the squeezing parameter is perceived as a higher temperature by the system. However, it is to be noted that even though the qualitative trends are similar for the one and two-qubit systems, quantitatively \textit{higher} power is being extracted in the present case (compare with fig. \ref{fig:PvsT}). 
As before, the red horizontal line is the locus of zero power, so that below this line, the machine ceases to act as an engine, i.e., power is injected into rather than being extracted from it.

\begin{figure}[h!]
\centering
    \begin{subfigure}{0.45\textwidth}
        \includegraphics[width=\textwidth]{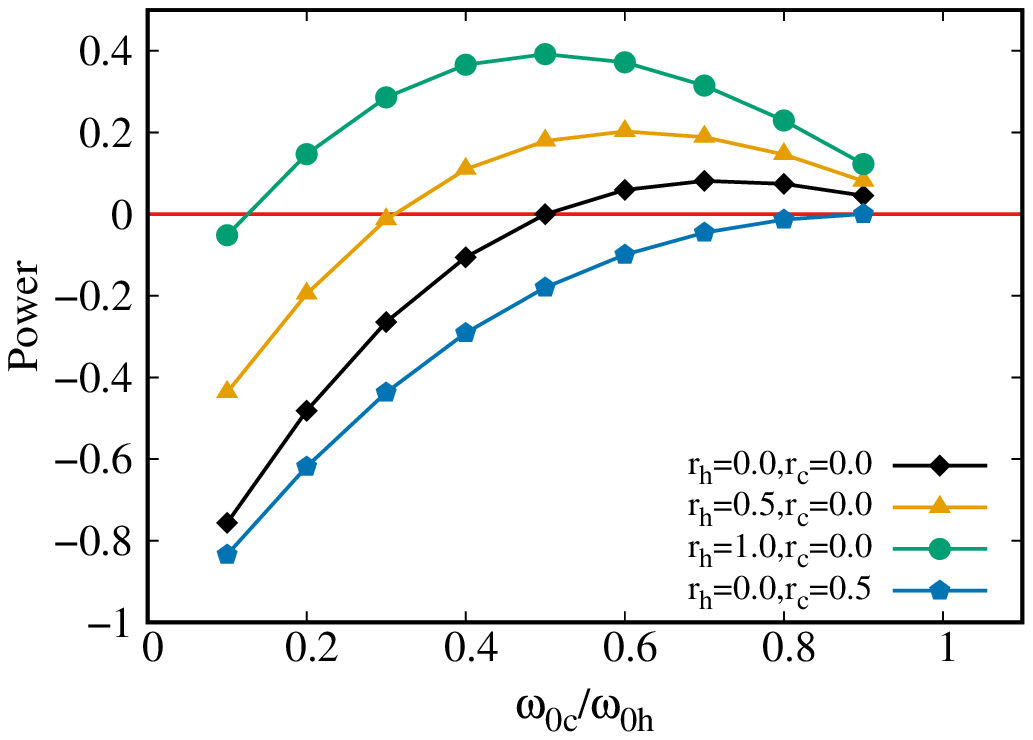}
        \caption{}
    \end{subfigure}
    \begin{subfigure}{0.45\textwidth}
        \includegraphics[width=\textwidth]{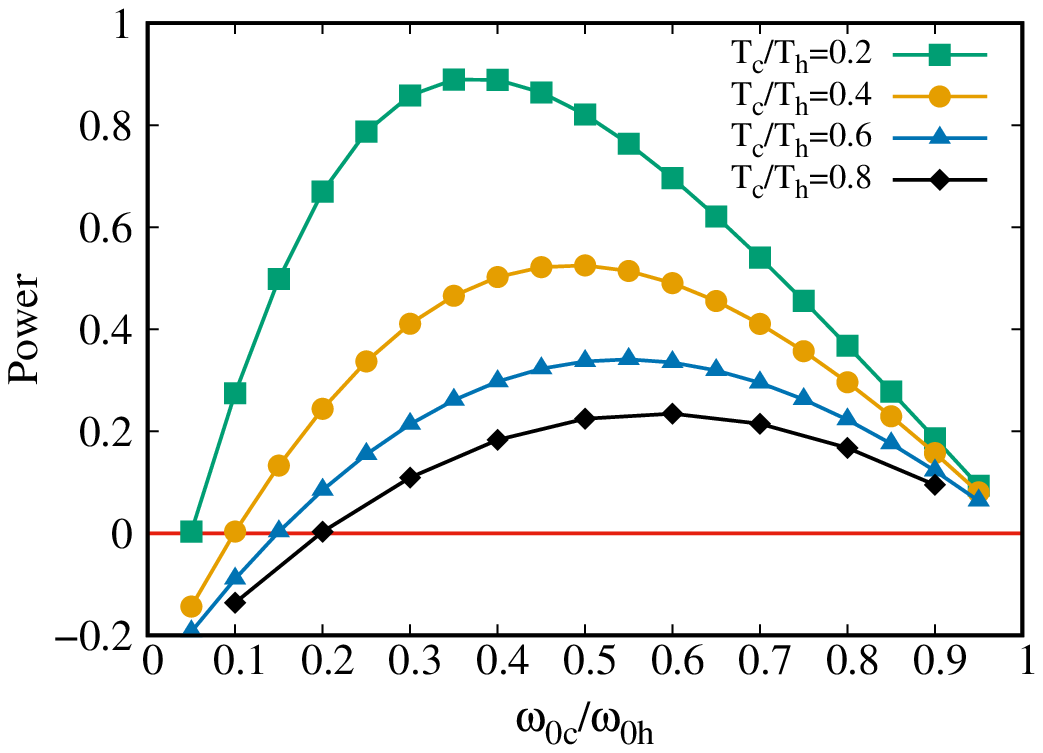}
        \caption{}
    \end{subfigure}

	\caption{(a) Variation of power of a TQE as a function of the frequency ratio $\omega_{0c}/\omega_{0h}$ for different squeezed parameters $r_h$ and $r_c$. Other parameters are: $T_h= 20$, $T_c= 10$, $r_{12}= 0.5$. (b) Power as a function of the frequency ratio $\omega_{0c}$/$\omega_{0h}$ for different values of bath temperature ratio $T_c/T_h$. The squeezing parameters are: $r_h= 1$, $r_c= 0$. Other parameters are same as in (a).}
	\label{fig:2qPvsFreq}
\end{figure}

 In fig. \ref{fig:2qPvsFreq}(a), the dependence of power on the frequency ratio $\omega_{0c}/\omega_{0h}$ has been shown, with $r_h$ and $r_c$ being the parameters that are varied. As the squeezing parameter $r_h$ is increased, the power is observed to increase. It is again observed that the power of the TQE is larger than that of the OQE (compare with fig. \ref{fig:PvsOmega}(a)).
 Fig. \ref{fig:2qPvsFreq} (b) shows the variation of power as a function of frequency ratio, but this time the temperature ratio $T_c/T_h$ is the parameter that is being varied, keeping the values of squeezing parameters fixed (compare with fig. \ref{fig:PvsOmega}(b)). In agreement with our intuitions, the curves of higher power are obtained when the temperature ratio is smaller, i.e., the difference between the bath temperatures is higher.

\begin{figure}[h!]
\centering
	\includegraphics[width=0.5\textwidth]{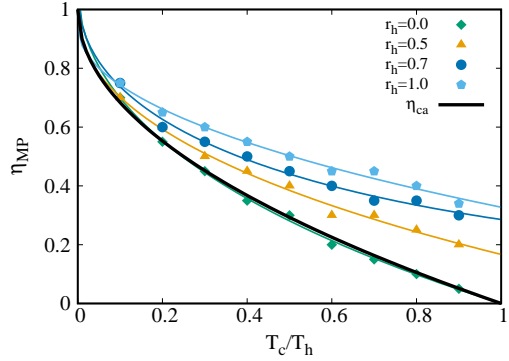}\\
	\caption{Efficiency at maximum power as a function of the temperature ratio $T_c/T_h$ for different squeezed parameters $r_h$. Other parameters are: $r_c= 0$, $r_{12}$= 0.5.}
	\label{fig:eta_mp}
\end{figure}

 \paragraph{Efficiency at maximum power:} The relationship between  $\eta_{MP}$ and the temperature ratio $T_c/T_h$ has been studied in fig. \ref{fig:eta_mp}. The method for plotting efficiency at maximum power is the same as that described for a single qubit. We observe that  $\eta_{MP}$ decreases as $T_c\to T_h$. 
 When all squeezing parameters are zero, the curve should converge to Eq. \eqref{eq:EtaCA}, which can be observed to be satisfied by the corresponding curve (green solid line/solid diamond symbols). Even though this is strictly true in the linear response regime, we find that the above form holds particularly well for the TQE even when $T_c\ll T_h$.

\begin{figure}[h!]
     \begin{center}
	\includegraphics[width=0.65\textwidth]{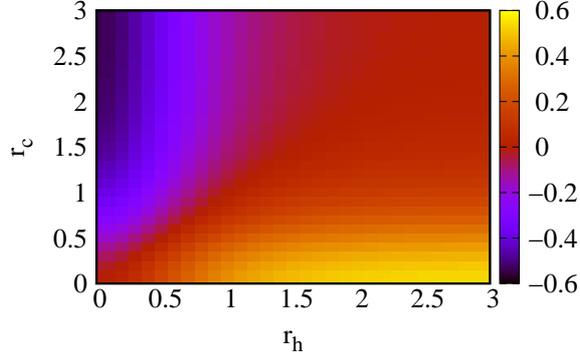}
	\caption{Phase plot of squeezing parameter $r_h$, $r_c$ and power of engine. The color coding provides the value of power. Other parameter used are: $\omega_{0c}$=10, $\omega_{0h}$=20, $T_c$=10, $T_h=20$, $r_{12}=0.5$.}
	\label{fig:denp}
	\setlength{\belowcaptionskip}{1pt}
	\end{center}
    \end{figure}

    \paragraph{} Fig. \ref{fig:denp} depicts the phase map between the output power of the TQE and the squeezing parameters of the bath. 
    It is to be noted that the machine generally works in the engine mode when $r_h>r_c$, since in the opposite regime the cold bath may be perceived by the system as the one with the higher effective temperature. Just like in the case of OQE, the machine may be made to act either as an engine or as a refrigerator by tuning the squeezing parameters.
    
  \paragraph{Effect of independent and collective decoherence:} 
  
\begin{figure}[h!]
    \centering
    \begin{subfigure}{0.48\textwidth}
   \includegraphics[width=\textwidth]{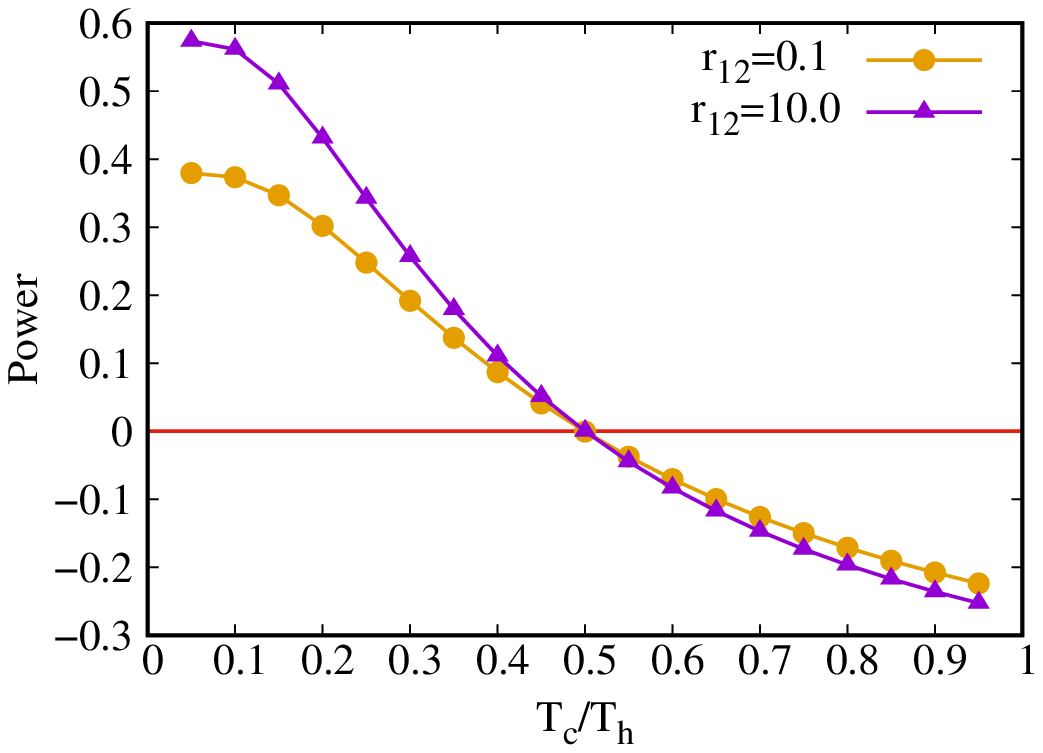}
	\caption{}  
	\end{subfigure}
    \begin{subfigure}{0.48\textwidth}
    \includegraphics[width=\textwidth]{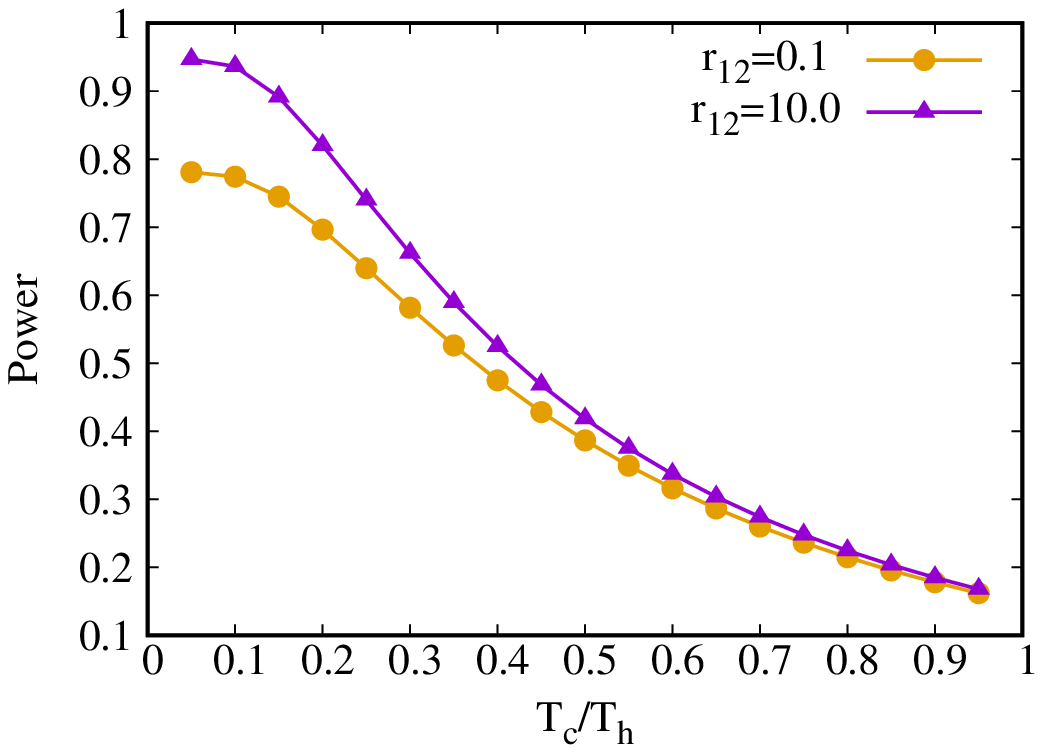}
    \caption{}
   \end{subfigure}
	\caption{Variation of Engine power as the function of $T_c/T_h$ at different $r_{12}$. Other parameters are $r_h=0$,  $r_c$= 0,  $\omega_{0h}$= 20, $\omega_{0c}$= 10. (b) Same plots when $r_h=1$.}
 	\label{fig:Eta_pow_r12_2q}
    \end{figure}

  We now study the effect of proximity of the two spins on the engine output (see fig. \ref{fig:Eta_pow_r12_2q}), given by the redefined variable $r_{12}=\bm{k_0\cdot r_{12}}$. Note that this is a new parameter that is relevant for the TQE, but is absent in the case of OQE.
  We may recall that $r_{12}\ll 1$ and $r_{12}> 1$ belong to the regimes of collective and individual decoherence, respectively. 
  A small value of $r_{12}$ (collective decoherence) would imply slower decay of the system to equilibrium, due to the reduced number of channels available for decay \cite{Ficek2002}, as compared to the scenario where $r_{12}$ is in the independent decoherence regime. Due to the faster decay in the latter regime, the process remains closer to an equilibrium one than in the former regime,
  thereby resulting in increased power output.
  Figure \ref{fig:Eta_pow_r12_2q}(a) corresponds to unsqueezed baths,  while fig. \ref{fig:Eta_pow_r12_2q}(b) corresponds to the presence of a squeezed hot bath. In both cases, the individual decoherence regime is found to yield a higher power than in the case of collective decoherence, although quantitatively the extracted powers in both cases are higher in presence of squeezing. It is also interesting to note that in presence of normal thermal baths, the system does not behave as an engine (yielding negative power) for a significant range of values of the temperature ratio. In contrast, in presence of squeezing the system remains an engine throughout the range.

\begin{figure}[h!]
 \centering
 	\begin{subfigure}{0.48\textwidth}
	\includegraphics[width=\textwidth]{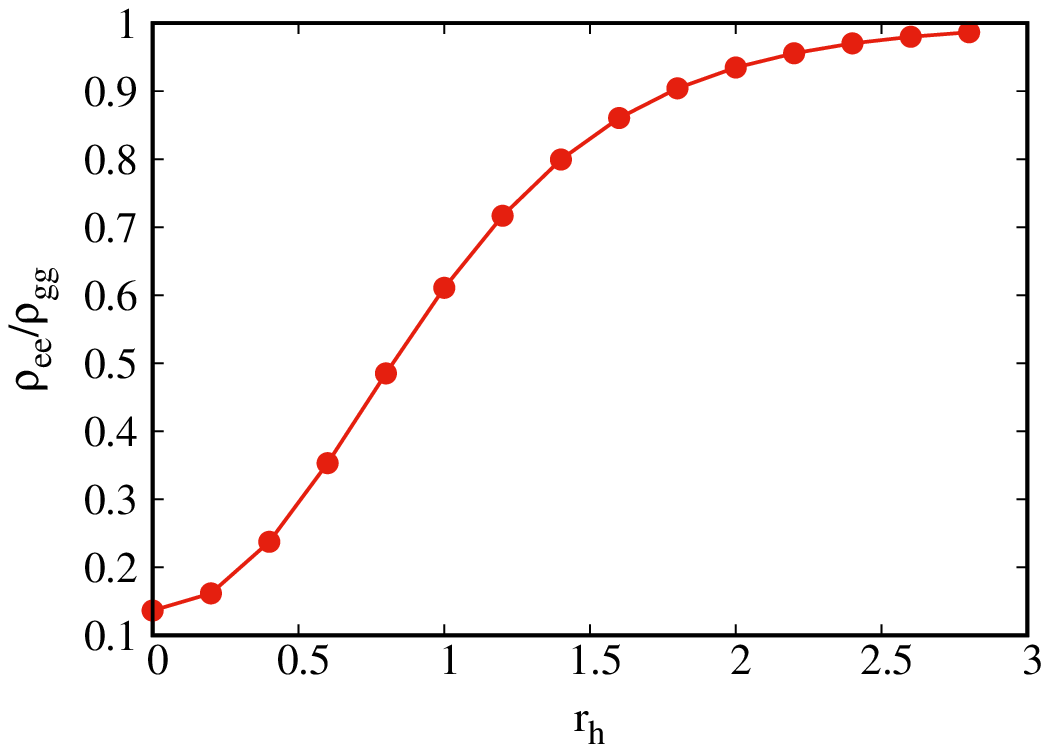}
	\caption{}
\end{subfigure}
  \begin{subfigure}{0.48\textwidth}
     \includegraphics[width=\textwidth]{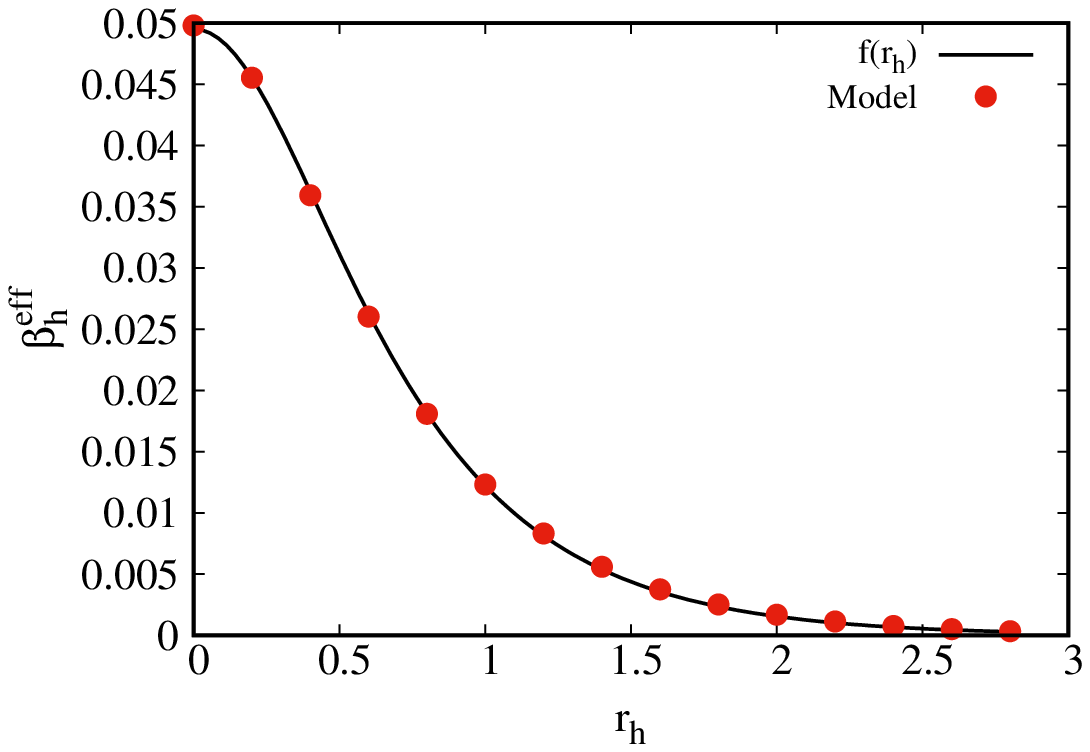}
     \caption{}
 \end{subfigure}
   \begin{subfigure}{0.48\textwidth}
     \includegraphics[width=\textwidth]{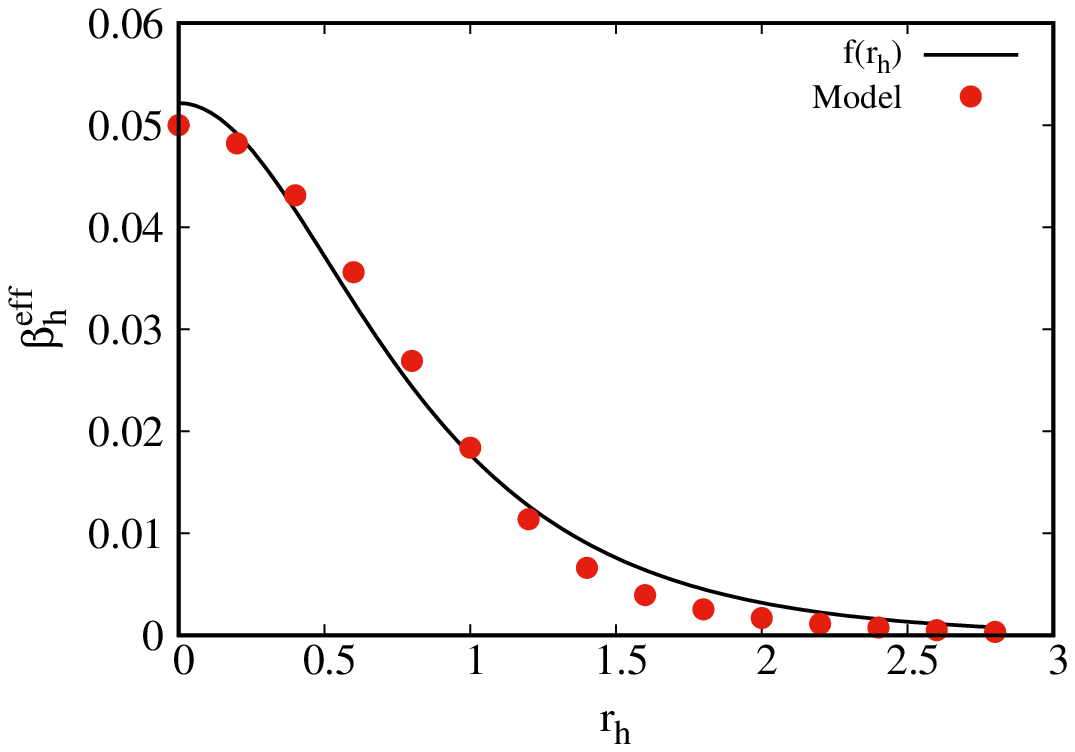}
     \caption{}
 \end{subfigure}
 \caption{(a)  Ratio of $\rho_{ee}$ and $\rho_{gg}$ as a function of $r_h$ for BC Stroke for the TQE.
 (b) Variation of effective inverse temperature with $r_h$ in state C. Here we have used the parameters: $\omega_{0h}=20$, $\omega_{0c}=10$, $T_h=20$ and $T_c=10$. Other parameters are: $r_{12}$= 0.5. The black solid line is given by the functional form $f(r_h)=a~\mathrm{sech}(br_h)$, where $a=0.0497, ~b=2.078$. (c) Same plot as (b), but far from stationary state of the system, when the system has evolved for a duration of 0.2 during stroke 2. The obtained parameters for the fitted curve are $a=0.052, ~b=1.742$.}  
    \label{fig:2D_Teff}
\end{figure}

\paragraph{Effective temperatures of two-qubit:} The dependence of the inverse effective temperature during stroke 2 of the engine on the squeezing parameters have been shown in Fig. \ref{fig:2D_Teff}, using a technique similar to the one used for OQE (see Sec. \ref{sec:OQE}). The variation of the population ratio (Fig. \ref{fig:2D_Teff}(a)) and of $\beta_h^{\rm eff}$ with $r_h$ (Fig. \ref{fig:2D_Teff}(b)) are same as in the case of OQE. The solid line in the Fig. \ref{fig:2D_Teff}(b) is the function given by $f(r_h) = a~{\rm sech}(br_h)$. The coefficients $a$ and $b$ are as mentioned in the figure caption. It is observed that the values of $a$ and $b$ closely correspond to the values obtained for OQE, and further verifies the relation $\beta^{\rm eff}_h = \beta~ \mathrm{sech}(2r_h)$ (see \cite{Manzano2018}).
 However, as has been shown in Fig. \ref{fig:2D_Teff}(c), while the system is out of its stationary state (i.e., during the relaxation phase at the beginning of stroke 2), the form of effective inverse temperature does not strictly follow the functional form $a~\mathrm{sech}(b r_h)$.
 Here, the data has been taken for the time instant $t=2.2$, which corresponds to the evolution of the system from state B towards state C, just for a duration of 0.2. The best fit obtained using the same form yields a curve that does not offer a very good fit. The final parameters obtained are $a=0.052$ and $b=1.742$, again showing that the form of the curve,  is both qualitatively and quantitatively different from the one that we obtain when the system has already relaxed to a stationary state. The sharp contrast of this behaviour with that of the OQE is to be noted; see discussion on fig. \ref{fig:Teff}(b).

\section{Conclusions} \label{sec:conclusions}

In this work, we have performed extensive numerical study of the thermodynamics of one and two-qubit Otto heat engines, in the nonequilibrium regime and in the presence of squeezed thermal reservoirs.  
The efficiency depends only on the ratio of the energy level spacing of the system (during the respective isochoric steps), and therefore is independent of squeezing parameters. This is to be contrasted with the power extracted from the engine, which shows significant sensitivity to the values of the squeezing parameters. We  study the variation of the output powers of the one-qubit engine (OQE) and two-qubit engine (TQE),
as a function  of the ratio of the bath temperatures, and of the ratio between the maximum and the minimum energy level spacings.
We also analyze the dependence of the efficiency at maximum power on the temperature ratio. We find that the Curzon-Ahlborn relation, Eq. \eqref{eq:EtaCA}, is valid for the OQE when $T_c\to T_h$  (expected linear response regime), whereas the relation remains closely valid for all values of $T_c/T_h$ for the TQE. We demonstrate that the inverse temperatures follow the same form as for a system at equilibrium with a squeezed heat bath, so that effective temperatures of the system grow exponentially with the squeezing parameters $r_h$ and $r_c$. This exponential growth shows different exponents when the system is at or away from its stationary state, for the TQE. For OQE, however, there is no perceptible difference between the two cases. The exponential growth gives us a good handle over the mode in which to operate our system, be it the engine mode or refrigerator mode. This is because the changed effective temperatures lead generally to an increased power output when $r_h>r_c$, and a smaller/negative output when $r_c>r_h$. 
\paragraph{}
The general observation is the enhanced power output of TQE as compared to the OQE. However, depending on the proximity of the two qubits in the TQE, the power may vary, even though it remains higher than the output from OQE. If the two qubits are in the collective decoherence regime ($r_{12}\ll 1$), the extracted power is typically smaller than in the individual decoherence regime ($r_{12}>1$). The variations of extracted power  on the squeezing parameters of the heat baths have been shown in a phase diagram, indicating the different modes (engine/refrigerator) of the system. We find that in the region $r_h>r_c$, the system generally behaves as an engine.
It would be an interesting and useful exercise to check how this general distinction between the one and two-qubit systems gets extended when we incorporate more qubits. 
 This is pertinent, as in the context of simulating general bosonic systems, harmonic oscillator eigenstates have been shown to be mapped to qubits \cite{somma2003quantum}. In the  present context, this could reveal interesting physics.
Making the squeezing parameters time-dependent may add to the richness of the phase plots, and this interplay between coherence and squeezing may be further exploited to unearth more applications of quantum heat engines.

\section*{Acknowledgements}

SL thanks DST, SERB for funding (grant number ECR/002607/2017). AK  thanks R. Ray for several  stimulating discussions. 
SB acknowledges support from Interdisciplinary Cyber Physical Systems (ICPS) programme of the Department of Science and Technology (DST), India, Grant No.: DST/ICPS/QuST/Theme-1/2019/13.
SB also acknowledges support from the Interdisciplinary Research Platform (IDRP)on Quantum Information and Computation (QIC) at IIT Jodhpur.

\appendix
\section{Dependence of efficiency on the system parameters}
\label{app:eta}

The average work done $\av{W_{AB}}$ from point A to B (see fig. \ref{fig:SchematicOtto}) is given by
\begin{align}
    \av{W_{AB}}=\mbox{Tr}[\rho_0\{H(\omega_c)-H(\omega_h)\}],
\end{align}
where $\rho_0=\rho_A=\rho_B$ is the density operator at the points $A$ and $B$ (which are same in the respective energy bases).
The average heat $\av{Q_{BC}}$ absorbed in the system from B to C is 
\begin{align}
    \av{Q_{BC}}=\mbox{Tr}[(\rho_1-\rho_0)H(\omega_h)],
\end{align}
$\rho_1$ being the density operator in state C.
Similarly,
\begin{align}
\av{W_{CD}} &= \mbox{Tr}[\rho_1(H(\omega_h)-H(\omega_c))],  \\
\av{Q_{DA}} &=\mbox{Tr}[(\rho_0-\rho_1)H(\omega_c)].
\end{align}
Thus, the total work extracted is
\begin{align}
\av{ W_{\rm tot} }= \av{W_{AB} + W_{CD}} = \mbox{Tr}[(\rho_0-\rho_1)(H(\omega_c)-H(\omega_h))]. \nonumber
\end{align}

 The efficiency of the engine is given by 
 \begin{align}
   \eta &=\frac{\av{W_{\rm tot}}}{\av{Q_{BC} } }\nonumber\\
	&=1-\frac{ {\rm Tr} [\Delta\rho H(\omega_c)]}{ {\rm Tr} [\Delta\rho H(\omega_h)]},
	\label{eq:eta_appendix}
 \end{align}
 where $\Delta\rho \equiv \rho_1-\rho_0$.
  Now let
  \begin{align}
      \Delta\rho &=
\begin{pmatrix}
  a & 0\\
 0 & b
 \end{pmatrix}.
 \end{align}
 We also know,
 \begin{align}
 H(\omega_c) &=
\begin{pmatrix}
 \omega_c/2 & 0\\
 0 & -\omega_c/2
\end{pmatrix},\nonumber\\
H(\omega_h) &=
\begin{pmatrix}
 \omega_h/2 & 0\\
 0 & -\omega_h/2
\end{pmatrix}.
\end{align}
Plugging these relations into Eq. \eqref{eq:eta_appendix}, the efficiency is obtained as:
  \begin{align}
 \eta &= 1-\frac{\omega_c}{\omega_h}.
\end{align}
A similar derivation can be done for the TQE, where Hamiltonian is used from Eq. \eqref{eq:two_hamiltonian}. The efficiency is again found to be independent of squeezing parameter $r$, and is given by 
 \begin{align}
 \eta &= 1-\frac{(a-d)\omega_{0c}+\Omega_{12}(b-c)}{(a-d)\omega_{0h}+\Omega_{12}(b-c)},
\end{align}
where we have used $\Delta\rho\equiv {\rm diag}(a,b,c,d)$.

\section{Solution of  the two qubit master Equation}
\label{app:twoQubit}

The following sixteen differential equations result from the solution of the density matrix Eq.\eqref{eq:two_master} which are listed below (see \cite{Banerjee2010b}) :-
\begin{align}
 \dot{\rho}_{ee}(t) &= -2\Gamma(\Tilde{N}+1)\rho_{ee}(t)+\Tilde{N}\{(\Gamma+\Gamma_{12})\rho_{ss}(t)+(\Gamma-\Gamma_{12})\rho_{aa}(t)\}+\Gamma_{12}|{\Tilde{M}|\rho_u(t)};\nonumber\\
&\text{where}\hspace{0.1 cm}\rho_u(t) = e^{i\phi}\rho_{ge}(t)+h.c.;\nonumber\\
\nonumber\\
\dot{\rho}_{es}(t) & = -i((\omega_0-\Omega_{12}))\rho_{es}(t)-\frac{1}{2}\{(3\Gamma+\Gamma_{12})+2\Tilde{N}(2\Gamma+\Gamma_{12})\}\rho_{es}(t)\nonumber\\
&\hspace{2.5 cm}+\Tilde{N}(\Gamma+\Gamma_{12})\rho_{sg}(t)+\Tilde{M}\Gamma_{12}\rho_{gs}(t)-\Tilde{M(\Gamma+\Gamma_{12})\rho_{se}(t)};\nonumber\\
\nonumber\\
 \dot{\rho}_{ea}(t) &= -i((\omega_0+\Omega_{12}))\rho_{ea}(t)-\frac{1}{2}\{(3\Gamma-\Gamma_{12})+2\Tilde{N}{(2\Gamma-\Gamma_{12}\}\rho_{ea}(t)}-\Tilde{N}(\Gamma-\Gamma_{12})\rho_{ag}(t)\nonumber\\
 &\hspace{5.5 cm}+\Tilde{M}\Gamma_{12}\rho_{ga}(t)-\Tilde{M}(\Gamma-\Gamma_{12})\rho_{ag}(t);\nonumber\\
 \nonumber\\
\dot{\rho}_{eg}(t) &= -2i\omega_0\rho_{eg}(t)+\Tilde{M}\Gamma_{12}-(2\Tilde{N}+1)+\Gamma\rho_{eg}(t)\nonumber\\
&\hspace{5.5 cm}-\Tilde{M}((\Gamma+2\Gamma_{12})\rho_{ss}(t)-(\Gamma-2\Gamma_{12})\rho_{aa}(t);\nonumber\\  
\nonumber\\
\dot{\rho}_{se}(t) &= \Dot{\rho}^*_{es}(t);\nonumber\\
\nonumber\\
\dot{\rho}_{ss}(t) &=-(\Gamma+\Gamma_{12})\{-\Tilde{N}+(1+3\Tilde{N})\rho_{ss}(t)-\rho_{ee}(t)-\Tilde{N}\rho_{ss}(t)+|\Tilde{M}|\rho_u(t)\};\nonumber\\
\nonumber\\
\dot{\rho}_{sa}(t) &= \dot{\rho}^*_{as}(t);\nonumber\\
\nonumber\\
 \dot{\rho}_{sg}(t) &= \dot{\rho}^*_{gs}(t);\nonumber\\
 \nonumber\\
\dot{\rho}_{ae}(t) &= \dot{\rho}^*_{ea}(t);\nonumber\\  
\nonumber\\
\dot{\rho}_{as}(t)  &= i2\Omega_{12}\rho_{as}(t)-\Gamma(1+2\Tilde{N})t\rho_{as}(t);\nonumber\\ 
&\text{The above equation can be trivially solved to yield};\nonumber\\
\rho_{as}(t)  &= \exp{[i2\Omega_{12}-\Gamma(1+2\Tilde{N})t]}\rho_{as}(0);\nonumber\\
\nonumber\\
\rho_{sa}(t)  &= \rho^*_{as};\nonumber\\
 \dot{\rho}_{aa}(t)  &= (\Gamma-\Gamma_{12})\{\Tilde{N}-(1+3\Tilde{N})\rho_{aa}(t)+\rho_{ee}(t)-\Tilde{N}\rho_{ss}(t)+|\Tilde{M}|\rho_u(t)\};\nonumber\\  
 \nonumber\\
\dot{\rho}_{ag}(t) &= \dot{\rho}^*_{ga}(t);\nonumber\\ 
\nonumber\\
\dot{\rho}_{ge}(t) &= 2i\omega_0\rho_{ge}(t)+\Tilde{M)*}\Gamma_{12}-(2\Tilde{N}+1)\nonumber\\
     &\hspace{4 cm}+\Gamma\rho_{ge}(t)-\Tilde{M}^*(\Gamma+2\Gamma_{12})\rho_{ss}(t)-(\Gamma-2\Gamma_{12})\rho_{aa}(t);\nonumber\\
     \nonumber\\
\dot{\rho}_{gs}(t) & = -i((\omega_0-\Omega_{12}))\rho_{gs}(t)-\frac{1}{2}\{(\Gamma+\Gamma_{12})+2\Tilde{N}((2\Gamma+\Gamma_{12})\}\nonumber\\
&\hspace{3 cm}\rho_{gs}(t)+(1+\Tilde{N})(\Gamma+\Gamma_{12})\rho_{se}(t)\Tilde{M}\Gamma_{12}\rho_{es}(t)-\Tilde{M(\Gamma+\Gamma_{12})\rho_{sg}(t)};\nonumber\\
\nonumber\\
\dot{\rho}_{ga}(t) &= i((\omega_0-\Omega_{12}))\rho_{ga}(t)-\frac{1}{2}\{(\Gamma-\Gamma_{12})+2\Tilde{N}{(2\Gamma-\Gamma_{12}}\}\nonumber\\
&\hspace{2 cm}\rho_{ga}(t)-(1+\Tilde{N})(\Gamma-\Gamma_{12})\rho_{ae}(t)+\Tilde{M}^*\Gamma_{12}\rho_{ea}(t)-\Tilde{M}^*(\Gamma-\Gamma_{12})\rho_{ag}(t);\nonumber\\
\nonumber\\
 \rho_{gg}(t) &= 1-\rho_{aa}(t)-\rho_{ss}(t)-\rho_{ee}(t).
 \end{align}
For an entanglement analysis, one can rotate the state from the dressed state basis back to the computational basis using a Hadamard transformation \cite{Banerjee2010b}.


\end{document}